\documentclass[aps,prb,twocolumn,groupedaddress,showpacs]{revtex4}

\usepackage{graphicx}
\usepackage{bm}

\begin{document}

\title{Huge magneto-crystalline anisotropy of x-ray linear dichroism
observed on Co/FeMn bilayers}

\author{W.\ Kuch}
\email{kuch@physik.fu-berlin.de}
\homepage{http://www.physik.fu-berlin.de/~ag-kuch}
\affiliation{
Freie Universit\"at Berlin, Institut f\"ur Experimentalphysik, 
Arnimallee 14, D-14195 Berlin, Germany}

\author{F.\ Offi}
\altaffiliation[Present address:]{
CNISM and Dipartimento di Fisica, Universit\`a Roma Tre, 
Via della Vasca Navale 84, I-00146 Roma, Italy.}

\author{L.\ I.\ Chelaru} 
\altaffiliation[Present address:]{
Universit\"at Duisburg--Essen,
Institut f\"ur Experimentelle Physik, Lotharstra{\ss}e 1, D-47057
Duisburg, Germany.}

\author{J.\ Wang}
\altaffiliation[Present address:]{
Department of Physics and HKU-CAS Joint Lab 
on New Materials,
The University of Hong Kong, Hong Kong, China.}

\author{K.\ Fukumoto}
\altaffiliation[Present address:]{
SPring-8, 1--1--1 Kouto, Sayo-cho, Sayo-gun, Hyogo 679-5198, Japan.}

\author{M.\ Kotsugi}
\altaffiliation[Present address:]{
Hiroshima Synchrotron Radiation 
Center, 2--313 Kagamiyama, Higashi-Hiroshima, 739-8526
Hiroshima, Japan.}

\author{J.\ Kirschner}
\affiliation{Max-Planck-Institut f\"ur Mikrostrukturphysik,
Weinberg 2, D-06120 Halle, Germany}

\author{J. Kune\v{s}}
\affiliation{Theoretical Physics III, Center for Electronic Correlations
and Magnetism, Institute of Physics,
University of Augsburg, D-86135 Augsburg, Germany.}
\affiliation{Institute of Physics,
Academy of Sciences of the Czech Republic, Cukrovarnick\'a 10,
162 53 Praha 6, Czech Republic.}

\date{23.03.2007}

\begin{abstract}
    We present an x-ray spectromicroscopic investigation of
    single-crystalline magnetic FeMn/Co
    bilayers on Cu(001), using X-ray magnetic circular (XMCD) and
    linear (XMLD) dichroism at the Co and Fe $L_3$ absorption
    edges
    in combination with photoelectron emission microscopy (PEEM). 
    Using the 
    magnetic coupling
    between the ferromagnetic Co
    layer and 
    the antiferromagnetic 
    FeMn layer we are able to produce magnetic
    domains with two different 
    crystallographic 
    orientations of the magnetic easy axis
    within the same sample at the same time.  We find a huge
    difference in the XMLD contrast between the two types of magnetic
    domains, which we discuss in terms of intrinsic magneto-crystalline
    anisotropy of XMLD of the Co layer.
    We also 
    demonstrate that due to the high sensitivity of the method, the
    small number of induced ferromagnetic Fe moments at the FeMn--Co
    interface is sufficient to obtain magnetic contrast from XMLD in a
    metallic system.

\end{abstract}

\pacs{75.70.Ak, 68.37.-d, 75.50.Ee}

\maketitle

\section{\label{intro}Introduction}

The recent interest in the magnetic coupling between
antiferromagnetic (AF) and ferromagnetic (FM) materials is motivated
by the quest for fundamental insight into the
phenomenon of exchange bias.\cite{Nogues99}  This effect, the discovery 
of which dates back to the 1950's,\cite{Meiklejohn56} manifests itself 
in a shift of the magnetization curve along the field axis. Nowadays 
the exchange bias effect is employed in
a variety of devices, such as
sensors or hard disk read heads,
based on magnetic thin films.\cite{Dieny91,Kools96} 

Only few methods can be used to study the spin 
structure of ultrathin antiferromagnetic films.  
While neutron diffraction and M\"ossbauer spectroscopy 
have been successfully employed to 
explore the spin structures of many bulk antiferromagnets already decades ago, 
both methods suffer from a lack of signal if 
films of a few atomic layers are to be investigated.  
X-ray magnetic linear dichroism (XMLD) in the soft x-ray absorption,  
on the other hand, is a method with sub-monolayer sensitivity.
XMLD refers to the difference between x-ray absorption spectra for
the plane of x-ray polarization aligned parallel and perpendicular
to the atomic moments.\cite{1126} 
By symmetry the XMLD signal does not depend on the orientation of 
magnetic moments but only on their axial alignment, and is thus suitable for 
the investigation of ferromagnetic as well as collinear antiferromagnetic spin 
structures,  
antiferromagnetic thin films in particular. Linear dichroic signal is also 
encountered 
in case of structural, not magnetic, reduction of the symmetry, 
as is commonly the case in the direction along the film normal
due to the presence of interfaces.\cite{1394,1558}  This can be eliminated if 
measurements are compared in which only the direction of the spin axis is varied, 
while the lattice geometry is fixed.  
In the following we will use the acronym ``XMLD'' for this situation only.

Unlike x-ray magnetic circular dichroism (XMCD), which in $3d$ 
metallic systems essentially measures  
integral quantities, namely the spin and orbital magnetic moments, 
the size and shape of XMLD depend  
also on the details of 
the electronic structure. Although integral sum rules have been put 
forward for XMLD,\cite{730} which relate the integral over the XMLD 
signal to the magneto-crystalline-anisotropy energy, this integral 
is usually much smaller than the amplitude of the plus--minus feature 
in the XMLD spectrum. Theoretical calculations predict that the latter may vary 
significantly with the crystallographic orientation of the magnetic 
moments.\cite{1295}
This is what we call the magneto-crystalline anisotropy of XMLD.

In this paper we study the magneto-crystalline anisotropy of XMLD
of a thin Co layer.  We use the 
AF--FM coupling
between the FeMn and Co layers to manipulate the orientation of the  
Co moments, namely the observation that Co moments in a Co/FeMn
bilayer align along $\langle110\rangle$ directions 
when in contact with a magnetically disordered (paramagnetic) FeMn layer (above 
its N\'eel 
temperature), while they prefer $\langle100\rangle$ directions
when the FeMn layer magnetically orders (antiferromagnetic).\cite{1213,1475}  
We take advantage of the fact that the N\'eel temperature
depends on the thickness of the FeMn layer.\cite{1246} Growing 
wedge-shaped samples we are thus able to study both 
$\langle110\rangle$-
and $\langle100\rangle$-oriented domains at the same time.

We use a photoelectron emission microscope 
(PEEM), as described in Refs.\ \onlinecite{705,1280,1315} 
for the microscopic
laterally resolved detection of the x-ray absorption cross section of
the FeMn/Co bilayers.  In combination with x-ray magnetic circular
dichroism (XMCD) in the soft x-ray absorption as a magnetic contrast
mechanism, PEEM is routinely used for the element-resolved observation
of magnetic domain patterns in multilayered
structures.\cite{1130,705,1301} XMLD can equally serve as the magnetic
contrast mechanism for PEEM if linearly polarized x-rays are used.  
Here, a
combined XMCD and XMLD spectromicroscopic investigation of
single-crystalline FeMn/Co bilayers on Cu(001) is presented.  We find
that the XMLD signal of the FM Co layer exhibits a strikingly
different behavior when in contact with a paramagnetic and
an antiferromagnetically ordered FeMn layer, while the XMCD contrast does
not differ appreciably.  We compare our observations to {\it ab
initio}\/ calculations of the $L_3$ XMLD in bulk fcc Co for different
crystallographic orientations of the magnetic moments.

Investigation of the influence of spin and electronic structure on the XMLD 
requires
single-crystalline samples with well characterized AF--FM interfaces.
Because of the small lattice
mismatch (0.4\%),\cite{867} Fe$_{50}$Mn$_{50}$ films (FeMn in the
following) on a Cu(001) single crystal are ideal candidates for
such investigations.  Epitaxial, virtually unstrained
FeMn films can be grown in a layer-by-layer mode by thermal deposition on Cu(001) 
at room
temperature.\cite{1246}  This provides an opportunity to study
the magnetic properties of an AF/FM system in single crystalline
FeMn/Co and Co/FeMn bilayers on Cu(001).\cite{1246,1213,1299}
Scanning tunneling microscopy revealed atomically smooth
interfaces with islands or vacancies of single atomic height.\cite{1445}
Based on XMCD-PEEM investigations of FM/FeMn/FM trilayers and on XMLD
spectroscopy experiments of Co/FeMn bilayers, we concluded previously that
a non-collinear three-dimensional spin structure is present
in the ultrathin FeMn layers, possibly similar to the so-called $3Q$
spin structure present in bulk FeMn.\cite{1384}
Combination of the Kerr magnetometry and XMCD-PEEM imaging showed
that the magnetic coupling across the interface is mediated
by step edges of single atom height,
while atomically flat areas do not contribute.\cite{KuchNatMat}

\section{\label{exp}Experiment}
All experiments were performed {\it in-situ}\/ in an ultrahigh vacuum
system with a base pressure below $10^{-8}$ Pa.  The disk-shaped
Cu(001) single crystal was cleaned by cycles of 1 keV argon ion
bombardment at 300 K and subsequent annealing at 873 K for 15 minutes.
The surface exhibited a sharp $(1 \times 1)$ low energy electron
diffraction pattern.  No contaminations were
detectable by Auger electron spectroscopy (AES).

The films were grown by thermal evaporation on the clean substrate at 
room temperature in zero external magnetic field.  Fe and Co
were evaporated by electron 
bombardment of high purity wires (99.99\% purity) of 2 mm
diameter, while a rod (99.5\% purity) of 4 mm diameter was
used for Mn.  Fe$_x$Mn$_{1-x}$ films of equiatomic composition ($x = 0.50 \pm 
0.02$)
were obtained by simultaneous evaporation of Fe and
Mn from two different sources.  During the deposition the pressure in the
chamber was kept below $5\times10^{-8}$ Pa. A typical evaporation
rate was 1 ML per minute.  The composition of the FeMn films was
estimated from the evaporation rates of the two sources, 
determined by medium energy electron diffraction (MEED), and
cross-checked by Auger electron spectroscopy peak ratios.  No
indication of segregation of Cu into or on top of the FeMn layers
was found.  The thickness of the films was determined by MEED, 
which shows pronounced layer-by-layer oscillations.\cite{1246}
The FeMn layer was grown in the form of small wedges of 200 $\mu$m 
width, using the method described in Ref.\ \onlinecite{1006}.

The experiments were performed at the UE56/2-PGM1 
helical undulator
beamline of the Berlin synchrotron radiation facility BESSY,
which can be set to deliver circularly polarized radiation of either 
helicity with a degree of circular polarization of about 80\%, or 
linear vertical or horizontal polarization of $>97\%$.\cite{1145} 
The set-up of the electrostatic PEEM was identical to that described in Refs.\
\onlinecite{705,1280,1315}.  
The light was incident at an angle of $30^\circ$ with respect to the sample 
surface.
Rotation of the sample about the 
surface normal allowed to take images for different x-ray azimuthal  
angles of incidence.  
Parameters were set to 
a lateral resolution of 350 nm, and a field of view of 60 $\mu$m.

The XMCD images represent a grayscale-coded absorption
asymmetry for opposite helicities of the circularly polarized x-rays at 
the $L_3$ absorption maximum (777.5 eV),
\begin{equation}
A_{XMCD}=\frac{I_+-I_-}{I_++I_-},
\end{equation}
i.e., the difference of absorption images acquired with opposite 
helicities divided by their sum.  For the quantitative analysis, 
background images acquired at lower photon energy (5 eV below 
the $L_3$ maximum) were subtracted.  

For XMLD, the maximum contrast was 
determined from a series of images acquired with 0.2 eV photon energy
step around the maximum of the Co $L_3$ absorption peak of a 6 ML Co/Cu(001) 
film, using $p$-polarized light. Maximum contrast was found 
between images taken at photon energies $E_1 = 776.5$ eV and $E_2 = 777.9$ eV.  
Since the acquisition time necessary to observe XMLD 
contrast at the Fe $L_3$ edge in FeMn/Co/Cu(001) bilayers was of the 
order of hours, no such photon energy sweeps were undertaken for the Fe $L_3$ 
edge 
(maximum at 707.5 eV); 
instead, the same relative photon energies as determined for the Co $L_3$ edge 
were tentatively used ($E_1 = 706.5$ eV and $E_2 = 707.9$ eV).
Images at the two photon energies were taken using $s$ and $p$ polarized 
x-rays. Because of the $30^{\circ}$ incidence, XMLD from in-plane magnetization 
is larger by a factor of $4/3$ for $s$ polarized excitation.  
Taking into account the opposite sign of the effect for the two
polarizations we used the following formula for the XMLD contrast
\begin{equation}
    A_{XMLD}=\frac{1}{2}\biggl[\frac{I_s(E_2)-I_s(E_1)}{I_s(E_2)+I_s(E_1)}-
    \frac{4}{3}\frac{I_p(E_2)-I_p(E_1)}{I_p(E_2)+I_p(E_1)}\biggr].
\end{equation}
The images thus reproduce quantitatively the difference between 
images taken at the higher photon energy minus images taken at the 
lower photon energy for $s$ polarized x-rays, divided by the sum of 
these images.

\section{\label{res}Results and discussion}

\begin{figure}
\includegraphics[width=0.69\columnwidth]{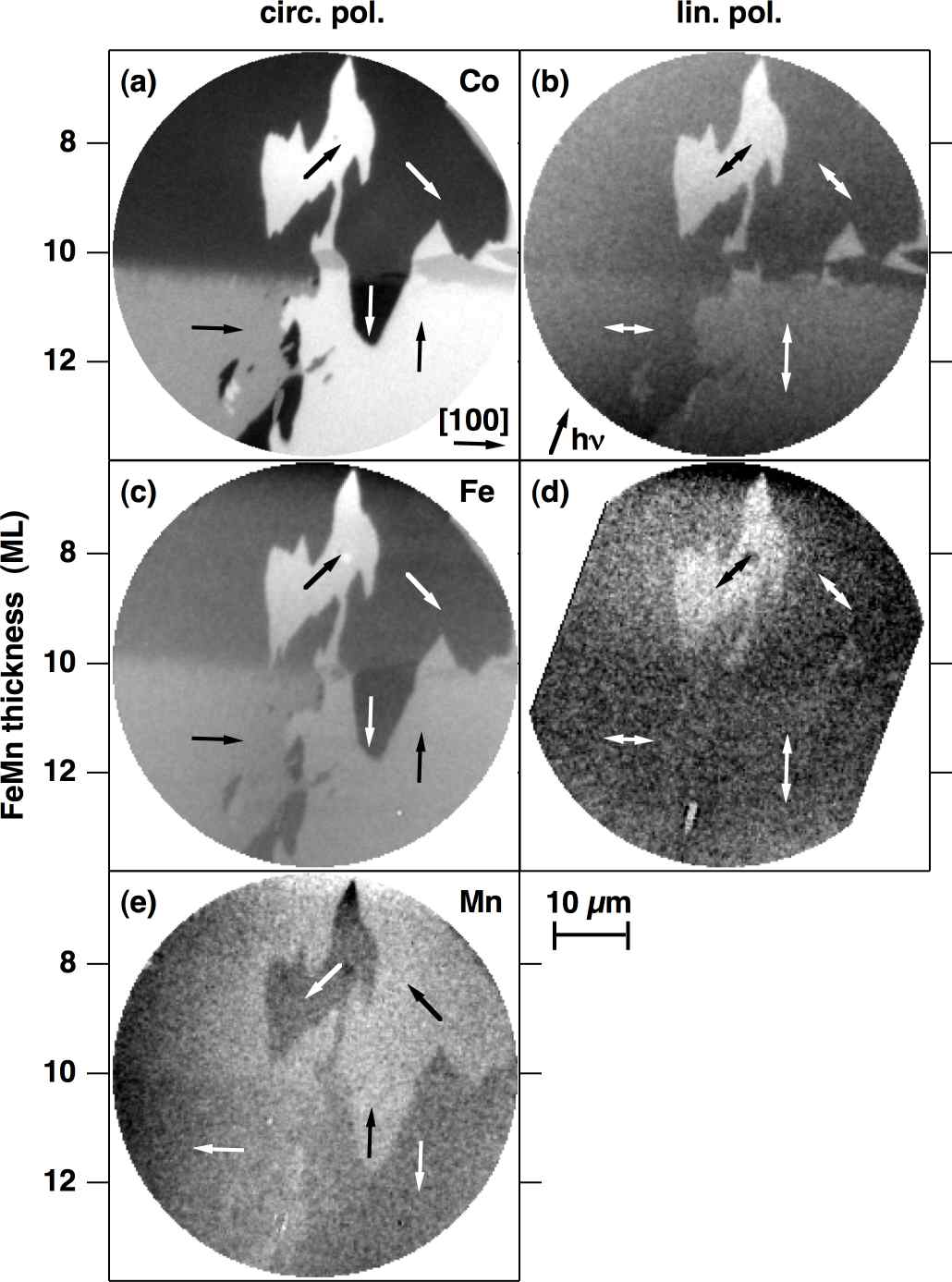}
\caption{Magnetic domain images of FeMn/Co/Cu(001) structure
obtained at the $L_3$ edges of 
(a), (b) Co, (c), (d) Fe, and (e) Mn.  
The thickness of the FeMn layer, increasing from the top to the
bottom of each image, is shown on the vertical axis.
Crystallographic orientation of the Cu substrate and 
azimuthal angle of incidence are shown in panels (a) and (b), 
respectively.  The left and right columns represent the XMCD and XMLD 
contrast, respectively. Arrows indicate the domain magnetization.
\label{elements}}
\end{figure}
This section is divided into three parts. First, we demonstrate the
performance of the spectromicroscopic domain imaging and present the
contrast obtained on the $L_3$ edges of Co, Fe and Mn. Next, we study
the dependence of the XMCD and XMLD contrast at the Co $L_3$ edge
on the azimuthal angle of incidence. Finally, we discuss the quantitative 
difference between the XMLD signal obtained from $\langle 110\rangle$ and
$\langle 100 \rangle$ domains. 

In  Fig.\ \ref{elements} typical XMCD and XMLD images obtained at the $L_3$
edge of Co are shown together with those obtained at the $L_3$ edges
of Fe and Mn.  A wedge-shaped FeMn/12 ML Co bilayer was used. 
The Co XMCD image (a) shows micron-sized magnetic domains. 
While the domain magnetization lies along $\langle 110 \rangle$
directions in the upper half of the image, only domains with
magnetization along $\langle 100 \rangle$ can be seen the lower half.
This behavior was observed previously,\cite{1213,1475} and is related to the fact
that for less than about 10 ML FeMn 
thickness
is paramagnetic at 
room temperature, while thicker FeMn layers develop AF order.\cite{1246}

Identical domain
patterns with strongly reduced XMCD contrast are observed at the Fe (c)
and Mn (e) $L_3$ edges.  The contrast arises due to the induced 
moments in the FeMn layer.\cite{1299} The Mn image is a negative
of the Fe and Co images,
indicative of an antiparallel orientation of the Mn moments 
with respect to the magnetization direction of the FM Co 
layer.  

Images of the same spot obtained with XMLD as magnetic contrast
mechanism are shown in the right column of Fig.\ \ref{elements}.  The magnetic
domain pattern is clearly visible at the Co $L_3$ edge (b). Note that
domains with opposite magnetization direction cannot be distinguished
by XMLD (compare the right lower parts of images (a) and (b)). 
The XMLD contrast at the Fe $L_3$ edge is much weaker than 
at the Co $L_3$ edge.  
Only after averaging over images with
about 170 minutes total acquisition time we were able to recognize at least the 
magnetic domains in the top part of the image. 

The images of 
Fig.\ \ref{elements} are presented on quite different grayscale 
ranges:  While the full contrast from saturated white to saturated 
black in the Co XMCD image (a) is 20\%, it is amplified to 8\% 
for the Fe (c) and 1.5\% for Mn (e) XMCD images, and amounts to 3\% 
in the Co XMLD image (b), and only 0.7\% in the Fe XMLD image (d).  

Our previously published XMLD spectra,\cite{1384} in which the XMLD
signal at the Fe $L_3$ edge is below the noise level, supported a
non-collinear antiferromagnetic 
arrangement of Fe spins in the AF FeMn layer.  In this
case only the small induced ferromagnetic moment in the FeMn layer,
the XMCD signal of which corresponds to about 30\% of the Fe atoms in
the interface atomic layer,\cite{1299} leads to an XMLD signal.
Although XMLD has been successfully applied in the past to image
antiferromagnetic domains in PEEM,\cite{813,882,976,1024} no attempt
was made on metallic antiferromagnets, and it is commonly believed
that the reduced crystal field
splitting of the electronic states in metals\cite{655,1203,1295} is 
prohibiting the use of XMLD for magnetic imaging.
Fig.\ \ref{elements} (d), however, shows that it is possible to image the 
XMLD signal, even of the comparably low number 
of the induced moments, by PEEM.  

Quantitative estimates support the interpretation of the contrast 
observed in Fig.\ \ref{elements} (d):
In the top part of the image, 
the Fe XMLD is about a factor of 6 weaker than 
the Co XMLD at the same position.  This is about the same ratio as 
between the respective XMCD contrasts in panels (a) and (c) at about 
6 ML FeMn thickness.  This size of induced ferromagnetic alignment is 
consistent with our earlier investigation of FeMn/Co 
bilayers.\cite{1299} 
Furthermore, the Fe XMLD originating from induced moments at the 
interface decreases with increasing total FeMn 
thickness, 
so that the expected Fe XMLD signal
would be within the noise of the measurement of the spectra of Ref.\ 
\onlinecite{1384}, which were taken for a 15 ML FeMn film.

\begin{figure}
\includegraphics[width=6.0cm]{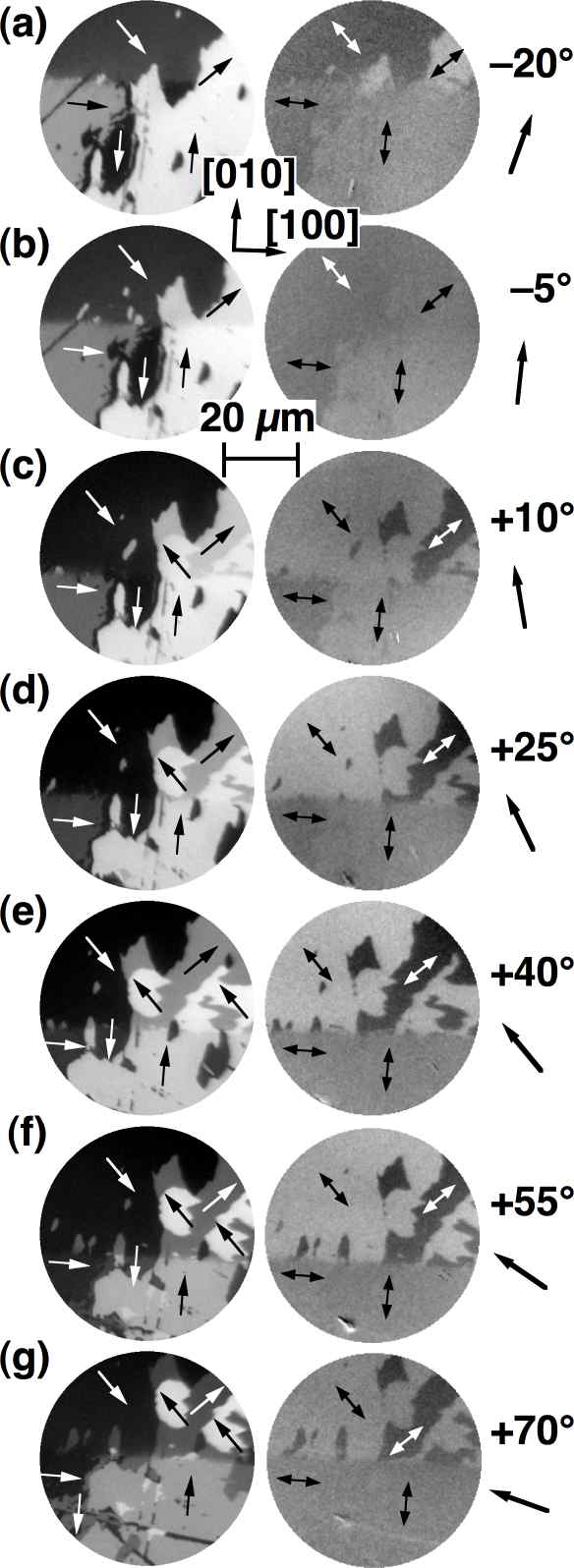}
\caption{Magnetic domain images of a FeMn/Co bilayer on Cu(001), 
acquired at the Co $L_3$ edge with 
circular polarization (left column) and linear polarization (right 
column).  Rows (a)--(g) correspond to different azimuthal angles of 
incidence, as labeled at the right hand side and indicated by 
arrows.  Local magnetization directions are indicated by 
arrows. The FeMn thickness increases from top to bottom 
in each image, from 7.2 to 14.7 in (a), gradually shifting to
from 9.4 to 16.9 ML in (g).
\label{angles}}
\end{figure}

Fig.\ \ref{angles} shows a series of magnetic domain images of the Co
layer from a sample in which a 0--25 ML wedge of FeMn was deposited on
top of a continuous film of 6 ML Co/Cu(001).  The left column shows
the XMCD contrast at the Co $L_3$ absorption maximum. 
The right column shows images of the same spot of the sample, 
acquired with linear polarization of the x-rays.  
As in Fig.\ \ref{elements}, the FeMn thickness increases from the top 
to the bottom of the images.  
Panels (a) through (g) show images obtained for different
azimuthal angles of incidence, indicated by an arrow at the right hand
side of each panel. Note that the field of view slightly shifted
due to readjustment of the sample.
The azimuth angle was read from a dial at
the sample holder with an accuracy of $1^{\circ}$.  $0^{\circ}$
corresponds to the nominal [010] direction of the Cu substrate;
however, as will be outlined below, the angular dependence of the
magnetic contrast indicates that the real [010] direction was at
$-2^{\circ}$ azimuth angle.  This deviation from the nominal direction
is within the accuracy with which the substrate could be oriented upon
mounting to the sample holder.

The data have been taken in the sequence from
(a) to (g).  Typical acquisition times were 4 minutes per
helicity for the circular polarization, and 20 minutes per
polarization direction and photon energy for the linear polarization.
Including the necessary sample manipulations, the time to obtain 
the data of Fig.\ \ref{angles} totalled 28 hours.
The time evolution of the domain pattern is clearly visible in the 
(a)--(g) series
from the shift of the transition line between paramagnetic and 
antiferromagnetic FeMn towards higher thicknesses. 
We attribute this to progressing contamination and possibly oxidation by
residual gas of the surface of the FeMn layer.  The Co layer, however,
which is investigated here, is protected against contamination by the
FeMn overlayer.

The right column of Fig.\ \ref{angles} shows images taken with linear 
x-ray polarization. The upper parts of the images 
clearly reveal identical domain pattern as seen with XMCD.
The behavior of the XMLD contrast follows the geometric expectations
including the reversal of contrast between panels (a) and (c), and
near vanishing of the contrast in panel (b) for the azimuthal angle
of incidence close to $45^{\circ}$ with respect to the magnetization.
The most prominent observation, and the main result of this work,
is the strong suppression of the XMLD contrast visible in the bottom
parts of the images which correspond to domains with $\langle 100 \rangle$
direction of magnetization.

\begin{figure}
\includegraphics[width=0.5\columnwidth]{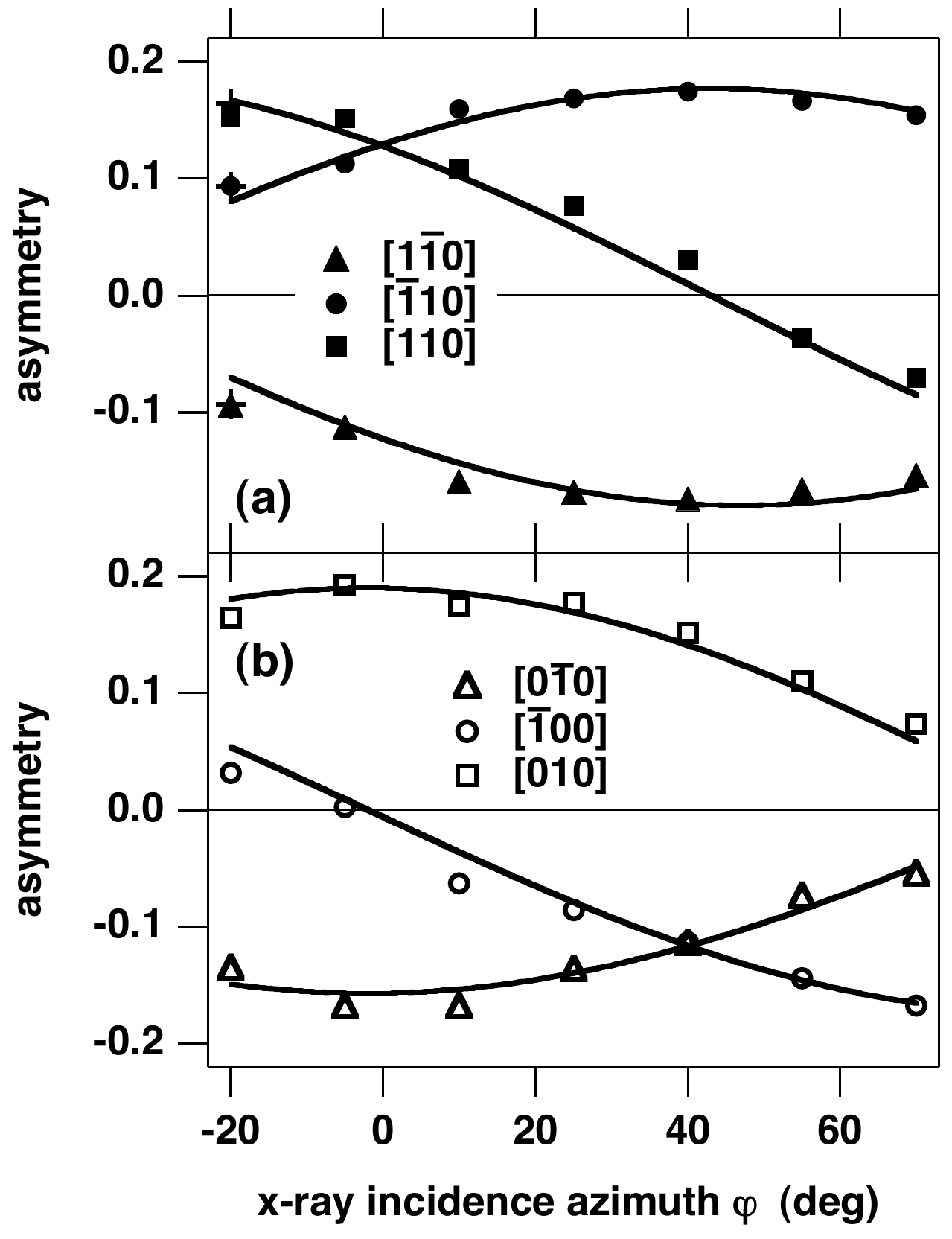}
\caption{
Angular dependence of the Co $L_3$ XMCD contrast: 
(a) the XMCD signal of three domains 
with magnetization directions $\lbrack 
1 \bar{1} 0 \rbrack$, $\lbrack 
\bar{1} 1 0 \rbrack$, and $\lbrack 
1 1 0 \rbrack$ represented by solid triangles, circles, and 
squares, respectively (the crosses at $-20^{\circ}$ are measurements 
from a Co/Cu(001) reference film without FeMn layer),
(b) the XMCD signal of three domains  
with magnetization directions $\lbrack 
0 \bar{1} 0 \rbrack$, $\lbrack 
\bar{1} 0 0 \rbrack$, and $\lbrack 
0 1 0 \rbrack$ represented by open triangles, circles, and 
squares, respectively.  
The solid lines are the result of simultaneous $\sin(\varphi)$ 
fits to the data.
\label{circ_angle}}
\end{figure}

In the following we discuss the angular dependence of XMCD and XMLD
contrast in detail. In Fig.\ \ref{circ_angle} we show
the Co XMCD contrast as a function of the azimuthal angle of incidence
obtained from the data of Fig.\ \ref{angles}.
Fig.\ \ref{circ_angle} (a) shows the contrast of the domains in the 
upper part of the images, 
where the FeMn is paramagnetic and Co domains are magnetized along 
$\langle 110 \rangle$ directions.  Panel (b) presents  
the contrast from the lower part of the images, where the 
FeMn is antiferromagnetic and Co domains align
along $\langle 100 \rangle$ directions. The lines correspond to 
a sine fit.
The fit reveals a $-2^{\circ}$ phase shift, which 
can be attributed to the inaccuracy of the sample 
mounting. The angular dependence of the 
XMCD contrast of the individual domains confirms very nicely the 
assignment of magnetization directions in 
Fig.\ \ref{angles}.  A small vertical offset may 
be attributed to instrumental asymmetries, as for example different 
intensities of the two helicities.  Importantly, the amplitude of the XMCD 
contrast is only about 5\% lower in panel (b) than in panel (a).  The 
latter is equal to the contrast of Co domains in a Co/Cu(001) 
reference sample without FeMn layer, which is indicated in Fig.\ 
\ref{circ_angle} (a) by crosses at $-20^{\circ}$ incidence azimuth.

\begin{figure}
\includegraphics[width=0.5\columnwidth]{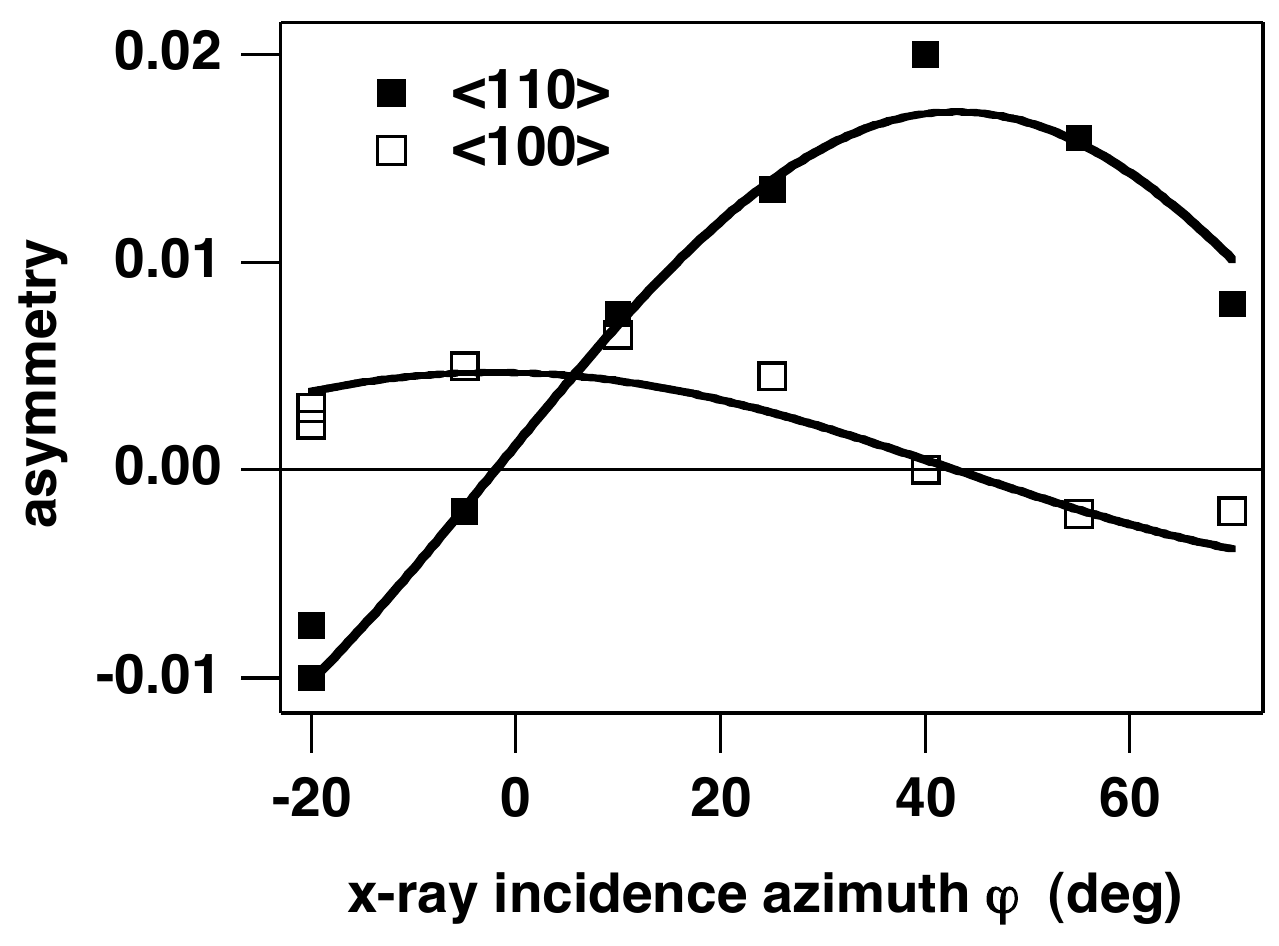}
\caption{
Angular dependence of the Co $L_3$ XMLD contrast. 
The contrast between Co magnetic domains 
with mutually perpendicular magnetization direction along $\langle 1 1 0 
\rangle$ directions is represented by solid symbols, the 
contrast between Co domains 
with perpendicular magnetization direction along $\langle 1 0 0 
\rangle$ directions is represented by open symbols.  The 
solid lines are the result of simultaneous $\sin(2\varphi)$ 
fits to the data.
\label{lin_angle}}
\end{figure}

Fig.\ \ref{lin_angle} shows the angular dependence of the Co XMLD 
contrast from data of Fig.\ \ref{angles}.  
Solid and open symbols represent the contrast between domains with 
mutually perpendicular magnetization direction in the 
regions where the FeMn layer is paramagnetic and antiferromagnetic, 
respectively.  The solid lines are the result of sine fits for 
which the phases were fixed at $+43^{\circ}$ and $-2^{\circ}$, 
respectively, using the result of the fits of 
Fig.\ \ref{circ_angle}.  Again, the data confirm the assignment of the 
magnetization axes in Fig.\ \ref{angles}. 
Based on the fits we are able to quantify the suppression
of the XMLD contrast in  $\langle 1 0 0
\rangle$ domains as compared to the $\langle 1 1 0
\rangle$ domains to be a factor of 0.28. Such a large effect of magnetization
direction on any physical quantity is rather unusual in $3d$ metals, 
which exhibit only a weak spin--orbit coupling. 
In order to prove that we see a genuine magneto-crystalline anisotropy
we next discuss the role of spin non-collinearity and make comparison
to other experimental data and theoretical calculations.
 
\begin{figure}
\includegraphics[width=0.5\columnwidth]{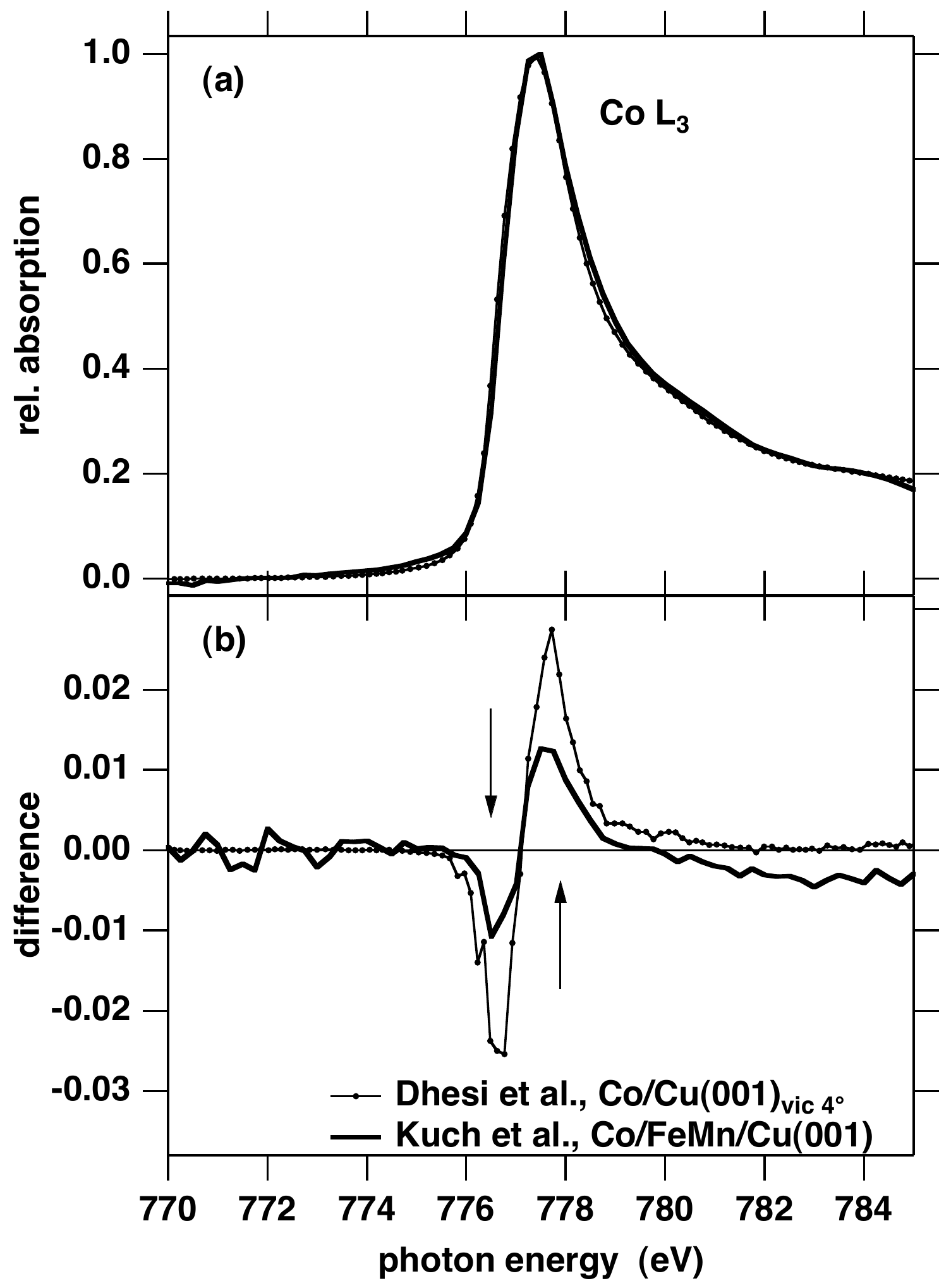}
\caption{
Comparison of different
experimental Co XMLD data from literature.  (a): Polarization-averaged 
absorption at the Co $L_3$ edge,
(b): Difference between absorption for parallel and
perpendicular x-ray polarization.  Markers and thin lines:
Fig.\ 1 of Ref.\
\onlinecite{1203}, for 6 ML Co on $4^{\circ}$ miscut Cu(001), 
magnetization along $\langle 110 \rangle$. 
Thick lines: Data from Fig.\ 4 of Ref.\
\onlinecite{1384}, for 15 ML FeMn/6 ML Co/Cu(001), magnetization 
along $\langle
100 \rangle$, scaled to the same absorption maximum.
The data of Ref.\ \onlinecite{1203} have been shifted in
energy by $-1.34$ eV for overlap of the absorption curves in (a).
\label{MLD_exp}}
\end{figure}

Besides changing the easy axis direction in the Co layer, magnetic ordering of the
FeMn may also induce a small non-collinearity of the Co moments.
Such a non-collinear fanning out of the FM moments in
Fe/MnF$_2$ bilayers as a consequence of the AF--FM coupling
was recently suggested on the basis of M\"ossbauer spectroscopy.\cite{1501}
We consider now if a similar scenario can explain our
XMLD data.
A distribution of the Co spins around a mean
direction would lead to a reduction of both the XMCD and XMLD signals
compared to the fully aligned case. While the reduction of the XMCD
signal is proportional to the reduction of the net moment, XMLD,
due to its different angular dependence, is more sensitive and depends
also on the distribution of the fanning angles.
In the extreme case of moments oriented at $45^{\circ}$ with respect
to the net magnetization, the XMLD 
would be reduced to zero, while the XMCD would still be  
at 71\% of its maximum value. 
Using the reduction factor of 0.95 of the net magnetization, 
obtained from XMCD contrast,
and assuming both binary 
(moments point at a fixed angle on either side of the net magnetization)
and normal distributions 
(Gaussian distribution of angles) 
of the fanning angle we arrive at a reduction factor of about 0.81 for the
XMLD contrast.  Therefore fanning of the Co moments due to the interaction
with the FeMn layer, if at all present, can explain only a small fraction
of the observed effect, which amounts to the reduction factor of 0.28.

\begin{figure}
\includegraphics[width=0.5\columnwidth]{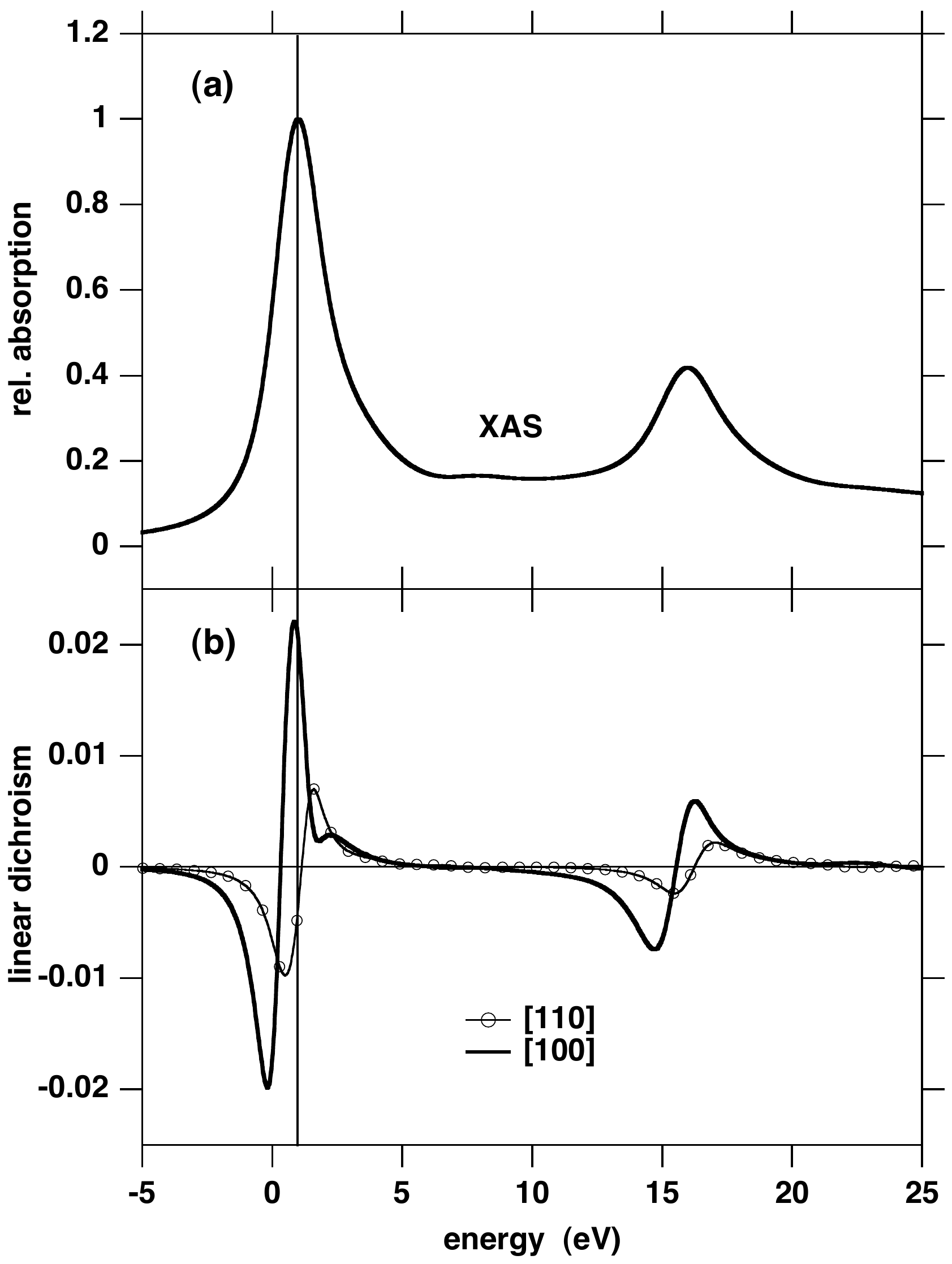}
\caption{
Calculated XMLD in the fcc Co for magnetization along two
different crystallographic directions.  (a): Polarization-averaged 
absorption spectrum for $\lbrack 100 \rbrack$ magnetization.
(b): XMLD difference for magnetization along $\lbrack 100 \rbrack$ (solid
line) and along $\lbrack 110 \rbrack$ (line and symbols).
The data for $\lbrack 100 \rbrack$ magnetization are
reproduced from Ref.\ \onlinecite{1295}. The energy scale is relative
to the absorption edge.
\label{MLD_theo}}
\end{figure}

Since the stability of the instrument is not sufficient to acquire  
series of microspectroscopic images for different photon energies, 
we use published data to compare experimental Co XMLD spectra 
for magnetization along $\lbrack 100 \rbrack$ and $\lbrack 110 
\rbrack$ directions.  The $\lbrack 100 \rbrack$ data are taken 
from Ref.\ \onlinecite{1384}, in which a 15 ML FeMn/6 ML 
Co bilayer on Cu(001) was measured. The spectra for Co magnetized along $\lbrack 
110 \rbrack$ 
direction are taken from the work of Dhesi {\it et al.}\/, in which 
6 ML Co on Cu(001), 
miscut by $4^{\circ}$, was measured.\cite{1203}  
The two spectra at the Co $L_3$ edge, rescaled to the same absorption 
maximum and shifted to the common position of the absorption edge,
are compared in Fig.\ \ref{MLD_exp}.  
Both spectra had been measured under similar conditions.\cite{footnote2}
Note that the peak-to-peak 
ratio of the XMLD spectra of Fig.\ \ref{MLD_exp} (b) cannot be compared
directly to the ratio of the XMLD asymmetry magnitudes of Fig.\ \ref{lin_angle},
which corresponds to the contrast between XMLD signal at the energies
marked by arrows in  Fig.\ \ref{MLD_exp}.\cite{footnote1}  
Although smaller than the asymmetry
anisotropy of 3.6 from Fig.\ \ref{lin_angle}, the 
ratio of peak-to-peak XMLD of 2.3 between the two curves
obtained from Fig.\ \ref{MLD_exp} still indicates
substantial magneto-crystalline anisotropy of the XMLD signal.

As mentioned in the Introduction, such a large anisotropy is 
unusual in $3d$ metals since the spin--orbit coupling is rather weak, e.g.\
the calculated magneto-crystalline anisotropy energy in bulk fcc Co
is only 2 $\mu$eV per atom. \cite{mae} Also the XMCD spectrum,
which depends essentially only on integral quantities, namely spin and
orbital moments, exhibits a very small anisotropy.\cite{1295} 
As pointed out by one of us and P.\ M.\ Oppeneer, the XMLD signal in metallic Co 
depends only weakly on the small valence band spin-orbit coupling.
The major contribution to XMLD comes from the exchange splitting
of the $2p$ levels ($\approx 1$ eV).\cite{1295} 
The magneto-crystalline anisotropy then arises from the fact that different
final $3d$ states are probed for different orientations of the
sample magnetization.  
 
To assess the feasibility of our experimental data, we used the calculated
XMLD spectrum 
of Ref.\ \onlinecite{1295} for the [100] direction
(what is referred to in Ref.\ \onlinecite{1295} as ``full calculation'')
and augmented these with 
equal
calculations for the [110] magnetization on the same system (see Fig.\ 
\ref{MLD_theo}).
In the calculations performed on bulk fcc Co a sizable anisotropy of XMLD
is found, however, the [100] exhibits larger XMLD magnitude contrary 
to the experiment. Before dismissing these results as a disagreement
a few remarks are in order. First, the calculations were done on bulk material
while the experiment is performed on a thin layer sandwiched by other materials, 
therefore a good quantitative agreement is unlikely. 
Second, we cannot judge the calculated anisotropy based on the present data only.
Note that due to a slight mutual shift of the calculated spectra, the [100]
contrast at the maximum amplitude of the [110] XMLD would be rather small.
Such a shift is not present in Fig.\ \ref{MLD_exp}, where [100] and [110]
spectra obtained on slightly different samples are compared.
Third, a possible non-collinearity of Co spins due to the presence of 
the AF FeMn layer would lead to local moments pointing neither along 
[110] nor fully along [100]. 
Taking these uncertainties into account we draw a modest, nevertheless 
non-trivial, conclusion that the
theory does not prohibit a magneto-crystalline
anisotropy of XMLD as large as observed in our experiment. 

\section{\label{sum} Conclusions}
We have presented a spectromicroscopic PEEM investigation 
of the magnetic domain pattern on Co/FeMn bilayers using XMCD 
and XMLD as the contrast mechanism. 
The sensitivity of the method allows to visualize even the tiny XMLD 
signal of the induced ferromagnetic moments in the FeMn layer.
We have found a factor of 3.6 difference in the XMLD contrast
between the Co $L_3$ signal from $\langle 110\rangle$ and 
$\langle 100\rangle$ domains in a {\it single} sample.
We argue that this huge difference is mainly due to an intrinsic
magneto-crystalline anisotropy of XMLD of the Co layer. 
Comparison of experimental XMLD spectra obtained from different
samples published previously and {\it ab initio} calculations
on bulk fcc Co suggest that such an anisotropy is indeed possible.

\begin{acknowledgments}
    We thank B.\ Zada and W.\ Mahler for technical assistance, and S.\
    S.\ Dhesi for providing the data from Ref.\ \onlinecite{1203}.
    Financial support by the German Minister for Education and
    Research (BMBF) under grant No.\ 05 SL8EF19 is gratefully
    acknowledged.  J.\ K. acknowledges the support by an Alexander
    von Humboldt Research Fellowship.
\end{acknowledgments}


\end{document}